\documentclass[aps,prd,showpacs,superscriptaddress,groupedaddress]{revtex4}  
\usepackage{graphicx}  
\usepackage{dcolumn}   
\usepackage{bm}        
\usepackage{amssymb}   

\begin{document}

\title{A unified approach to electron and neutrino elastic scattering off nuclei with an application to the study of the axial structure\\
 }

\author{O. Moreno}
\author{T. W. Donnelly}
\affiliation{Center for Theoretical Physics, Laboratory for Nuclear Science and Department of Physics, Massachusetts Institute of Technology, Cambridge, MA 02139, USA}
\date{\today}

\begin{abstract}
We show a relationship between elastic electron scattering observables and the elastic neutrino cross section that provides a straightforward determination of the latter from experimental data of the former and relates their uncertainties. An illustration of this procedure is presented using a Hartree-Fock mean field for the nuclear structure of a set of even-even nuclear targets, using the spectra of the neutrinos produced in pion decay at rest. We also analyze the prospects to measure the incoherent axial contribution to the neutrino elastic scattering in odd targets.
\end{abstract}


\maketitle

In lepton-nucleus elastic scattering the incident and the outgoing lepton is the same and its energy loss $\omega$ is transformed entirely into kinetic energy of the recoiling nuclear target; we denote the process as $(\nu,\nu)$ for neutrinos (of any flavor) and $(e,e)$ for charged leptons (again of any flavor, but electrons being of most experimental interest). Coherent scattering is a particular case of elastic scattering where all of the nucleons in the target contribute to the cross section through the vector Coulomb monopole isoscalar form factor of the nucleus, which is, unlike the rest of incoherent elastic form factors, proportional to the number of nucleons. Coherence applies for momentum transfers corresponding to nuclear-size wavelengths, $q\sim 160 \:A^{-1/3}$ MeV, and below; for larger values the Coulomb form factor decreases and the incoherent elastic form factors, when possible (see below), become comparable.

Elastically scattered charged leptons can be easily detected, but in the case of neutrinos the proposed observable is the recoil energy of the nuclear target through the ionization induced in the detector. Elastic neutrino scattering off nuclei can be exploited to determine electroweak parameters at very low momentum transfers, to test the universality of the weak interaction for charged and neutral leptons, or to estimate the escape rate of neutrinos created in a variety of stages of star evolution. These motivations support recent experimental proposals to measure neutrino elastic scattering, such as the neutrino program at SNS-ORNL \cite{sch06} and the analysis of sensitivity to this process of several neutrino and dark matter detectors \cite{and11}.

Parity-violating (PV) elastic electron scattering is another nuclear electroweak process that has drawn much attention recently. The usual observable is the parity-violating asymmetry, defined as the relative difference between the cross sections of electrons with spin projection parallel (same direction, $h=+1$) and antiparallel (opposite direction, $h=-1$) to their momentum:
\begin{equation}\label{asymmetry_sigmas}
\mathcal{A}_{(e,e)} = \frac{\left(\frac{d\sigma}{d\Omega}\right)^{h=+1} - \left(\frac{d\sigma}{d\Omega}\right)^{h=-1}}{\left(\frac{d\sigma}{d\Omega}\right)^{h=+1} + \left(\frac{d\sigma}{d\Omega}\right)^{h=-1}}
\end{equation}

Measurements of parity-violating elastic electron scattering off nuclei can be used for precise tests of the Standard Model (SM), including the evaluation of the weak mixing angle or of higher-order radiative corrections, as well as to determine the neutron radii of nuclei \cite{dds}, with implications to the neutron-rich matter equation of state and to the structure of neutron stars. Recent or planned experimental efforts such us PREX I and II, using $^{208}$Pb \cite{prex}, and CREX, using $^{48}$Ca \cite{crex}, have focused on the extraction of the neutron radii of the target nuclei with precisions as good as 1.2$\%$ in the PV observable. There has also been recent interest in relatively low-energy electron beams for studies of PV electron scattering, such as the MESA accelerator at Mainz \cite{MESA} or an upgraded version of the FEL at Jefferson Lab \cite{FEL}, aimed at tenths of percent precision in the PV measurements.

The dominant electron-nucleus scattering process is overwhelmingly an electromagnetic (EM) one and therefore parity-conserving (PC). On the other hand, the weak neutral current (WNC) is responsible for the parity violation in electron scattering, since it contains vector and axial components that behave differently under inversion of spatial coordinates, and it is also responsible for neutrino-nucleus scattering. The probabilities of PC electron, PV electron and neutrino scatterings follow approximately the ratio 1 : 3$\cdot$10$^{-4}$ $q^2$ : 3$\cdot$10$^{-10}$ $q^4$, with $q$ the characteristic momentum transfer of the process in GeV. In what follows we consider the exchange of a single gauge boson for each of the interactions involved: one $Z^0$, one photon, or one of each; we also neglect the distortion of the electron wave functions due to the nuclear Coulomb field, although in practice it is usually taken into account. These two conditions are known as plane wave Born approximation (PWBA).

The elastic neutrino cross section differential with respect to the outgoing neutrino solid angle can be written as
\begin{eqnarray}
\left( \frac{d\sigma}{d\Omega} \right)_{(\nu,\nu)} = \frac{1}{2\pi^2} \:G_F^2 \:\varepsilon'^2_{\nu} \:\cos^2(\theta/2) \:f_{rec}^{-1} \:\widetilde{{\mathcal R}} \;,
\label{neutrino_xs}
\end{eqnarray}
where $\widetilde{{\mathcal R}}$ stands for the square of the WNC matrix element of the scattering process, namely the contraction of the corresponding leptonic and hadronic tensors (see later for the normalization chosen).

It is also useful to express the cross section in a form that is differential with respect to the target recoil energy (equal to the energy transfer), related to the previous expression through a Jacobian,
\begin{eqnarray}
\left( \frac{d\sigma}{d\omega} \right)_{(\nu,\nu)} = J(\Omega,\omega) \:\left( \frac{d\sigma}{d\Omega} \right)_{(\nu,\nu)} \:,
\end{eqnarray}
which is given by
\begin{eqnarray}
J(\Omega,\omega) = \frac{d\Omega}{d\omega} = \frac{2\pi \:(M_A+\omega) \:f_{rec}}{k_{\nu}\:k_{\nu}'\:(1+\omega/M_A)}.
\end{eqnarray}
In these expressions $\varepsilon_{\nu}$ and $\varepsilon'_{\nu}$ are the initial and final neutrino energies, respectively ($k_{\nu}$ and $k'_{\nu}$ the corresponding momenta), $\theta$ is the neutrino scattering angle, $M_A$ is the target mass and $f_{rec}$ is a kinematic recoil factor.

The differential neutrino cross sections imply the detection of the recoiling energy or momentum (magnitude or direction) of the target with reasonable precision; if, on the contrary, the detectors have a large energy acceptance from a minimum value ($\omega_m$, given by the detector threshold), up to a maximum value ($\omega_M$, given by the specific kinematic conditions), what is actually measured is
\begin{equation}\label{integrated_nu_xs}
\sigma_{(\nu,\nu)}(\omega_m) = \int_{\omega_m}^{\omega_M} \left(\frac{d\sigma}{d\omega} \right)_{(\nu,\nu)} \:d\omega .
\end{equation}

The matrix element squared in Eq. (\ref{neutrino_xs}) particularized to coherent neutrino scattering is $\widetilde{{\mathcal R}}=\widetilde{{\mathcal R}}_{coh}$, with
\begin{eqnarray}
\widetilde{{\mathcal R}}_{coh} = V_{L} \:(\widetilde{F}^{VV,\:T=0}_{CC,\:J=0})^2 \;,
\label{R_coh}
\end{eqnarray}
where $\widetilde{F}^{VV,\:T=0}_{CC,\:J=0}$ is the WNC Coulomb monopole vector isoscalar form factor, normalized so that in the long wavelength limit (LWL), {\it i.e.}, as the momentum transfer goes to zero, it becomes
\begin{eqnarray}
\widetilde{F}^{VV,\:T=0}_{CC,\:J=0}(q\to 0) \to A\:\sin^2\theta_W \;,
\label{F_norm}
\end{eqnarray}
where $A$ is the target mass number and $\theta_W$ is the weak mixing angle, $\sin^2\theta_W\approx 0.23$. The same normalization for the full Coulomb monopole form factor (isoscalar plus isovector) in LWL yields the nuclear weak charge ${\mathcal Q}_W = Z \:\beta^p_V + N \:\beta^n_V$, where $\beta^p_V=0.5-2\sin^2\theta_W\approx 0.04$ and $\beta^n_V=-0.5$ are the proton and neutron WNC vector coupling constants, respectively \footnote{Note that different normalizations have been used in the literature, leading to different factors in Eq. (\ref{neutrino_xs}). The Coulomb form factor might contain an extra factor $1/\sqrt{4\pi}$ and the WNC coupling constants might contain an extra factor 2; for the latter we use the conventions in \cite{don79_pr}.}. The Rosenbluth factor $V_{L}$ in the extreme relativistic limit (ERL) is $V_{L}= \alpha^{\nu}\:(1-\omega^2/q^2)^2$ where $\alpha^{\nu} = \left[ (a^{\nu}_A)^2 + (a^{\nu}_V)^2) \right] / 2 $ is a combination of neutrino WNC coupling constants, with $(a^{\nu}_A)^2 = (a^{\nu}_V)^2=1$ in the SM.

\section{Relationship between electron and neutrino coherent cross sections}

The elastic electron cross section, the parity-violating asymmetry in elastic electron scattering and the elastic neutrino cross section for even-even nuclear targets fulfill the following relationship:
\begin{eqnarray}\label{relationship}
\left( \frac{d\sigma}{d\Omega} \right)_{(\nu,\nu)} = \mathcal{A}_{(e,e)}^2 \: \left( \frac{d\sigma}{d\Omega} \right)_{(e,e)} \;,
\end{eqnarray}
where the cross sections and the asymmetry are evaluated at the same kinematic conditions (incident momentum and scattering angle) and the ERL for the leptons has been assumed. An additional factor of WNC leptonic couplings, namely $\alpha^{\nu}/(a_A^e)^2$, has been particularized to its SM value of 1; we note in passing that the neutrino scattering on which we are focused is insensitive to the values of $a_V^{\nu}$ and $a_A^{\nu}$ independently, and therefore to the possible Majorana nature ($a_V^{\nu}= 0$) of the neutrinos.

The relationship in Eq. (\ref{relationship}) is valid for any neutrino flavor and for any charged lepton flavor (as long as the ERL still holds), and for leptons as well as for antileptons in any combination, always within PWBA. For non even-even, $J\neq 0$ targets the relationship is only an approximation, since other contributions to the elastic scattering arise beyond the vector Coulomb monopole $M_0$; they are, however, smaller than the coherent contribution by factors $\sim Z^2$ (EM case) or $\sim N^2$ (WNC case), and only two of them, the axial longitudinal dipole $L_1^A$ and the axial electric dipole $T_1^{el.\:A}$ survive in the limit $q\rightarrow 0$ (see discussion below) \cite{don79_pr}.

For even-even $N=Z$ nuclei and neglecting nucleon strangeness content (see \cite{gon13} for its current experimental status), or at low enough momentum transfers, Eq. (\ref{relationship}) takes on an even simpler form \cite{mor09_npa}:
\begin{eqnarray}\label{relationship_isoscalar}
\left( \frac{d\sigma}{d\Omega} \right)_{(\nu,\nu)}^{N=Z} = \kappa \:Q^4 \: \left( \frac{d\sigma}{d\Omega} \right)_{(e,e)}^{N=Z} \end{eqnarray}
with $\kappa=$ 6.84$\cdot 10^{-9}$ GeV$^{-4}$ and $Q^2$ the four-momentum transfer squared. Under these conditions the neutrino cross section is purely coherent except for a small contribution from isospin-breaking effects (mainly of Coulomb origin). As before, it has to be corrected for other incoherent contributions in $J\neq 0$ nuclei ($^{2}$H, $^{6}$Li, $^{10}$B, $^{14}$N), but in the $q\rightarrow 0$ limit the approximation is much better than in the general case: the axial contributions that would survive in this limit do not actually take part because the isoscalar WNC axial coupling constant is zero at tree level in the SM; only isospin-breaking effects, known to be very small, introduce isovector contributions.

To clarify the discussion on coherent and incoherent contributions to WNC elastic neutrino scattering, we list in Table \ref{multipoles} the multipole operators and responses involved in the process with their characteristic factors and typical sizes; as mentioned above, the coherent contribution carries an extra factor proportional to the mass number. In the cross-section each response enters squared and multiplied by the corresponding generalized Rosenbluth factor, that can further reduce the relative weight of each contribution with respect to the coherent one; these considerations will be important for the next section on the determination of the axial form factor.

\begin{table}[h!]
\setlength\extrarowheight{2pt}
\caption{\small Elastic vector and axial multipole operators ($M_J$ for Coulomb, $L_J$ for longitudinal, $T^{el.}_J$ for transverse electric and $T^{mag.}_J$ for transverse magnetic) and WNC responses (purely vector, purely axial and vector-axial interference) when no tensor second-class currents are present, ordered according to the size of their characteristic factor at leading order for small momentum transfers, $q \approx$ 50 MeV, and using a typical nuclear (Fermi) momentum scale $q_N \approx$ 250 MeV and nucleon mass $m_N\approx$ 1 GeV \cite{don79_pr}.\label{multipoles}} 
\begin{center}
\begin{tabular}{ccccccc}
\\
\hline
\hline
Characteristic  & $\qquad$ Typical $\qquad$  &  \multicolumn{2}{c}{Multipole operators } & \multicolumn{3}{c}{WNC responses} \\ \cline{3-7}
 factor & size &  $\quad$Vector$\quad$ &$\quad$ Axial$^{\:\dagger}$ $\quad$& $\quad$$\quad$Vector$\quad$ & $\quad$Axial$^{\:\dagger}$$\quad$ & Interference\\
\hline
1 & 1 & $M_0^{\;\;\ddagger}$ & $L_1^{A}, \:T_1^{el.A}$  & $\quad$$\widetilde{F}^{VV{\;\ddagger}}_{CC}$ & $\quad$$\widetilde{F}^{AA}_{LL},\:\widetilde{F}^{AA}_{T}$ $\quad$\\
$q/m_N$ & 1/20 & $T_1^{mag.}$ &  & $\quad$$\widetilde{F}^{VV}_{T}$ &   & $\widetilde{X}^{VA}_{T'}$  \\
$(q/q_N)^2$ & 1/25 & $M_2$ &  $L_3^{A},\:T_3^{el.A}$   \\
\end{tabular}
\end{center}
\begin{center}
\begin{tabular}{c}
\footnotesize{$\dagger$ Only isovector in the SM.} \\
\footnotesize{$\ddagger$ Additional coherence factor ($\propto A$) for the isoscalar part. CVC assumed, $L_0=(\omega/q)\:M_0$.}
\end{tabular}
\end{center}
\end{table}

The relationships in Eq. (\ref{relationship}) or (\ref{relationship_isoscalar}) can also relate neutrino and electron cross sections for different kinematic conditions but with the same energy transfer $\omega=\omega_{e}=\omega_{\nu}$ by introducing the following factor, valid in ERL:
\begin{eqnarray}
K = \frac{k_{e}^2\:(k_{\nu}-\omega)^2 \:[2\:k_{\nu}^2-\omega\:(2\:k_{\nu}+M_A)]}{k_{\nu}^2\:(k_{e}-\omega)^2 \:[2\:k_{e}^2-\omega\:(2\:k_{e}+M_A)]} \;.
\end{eqnarray}

The main practical application of Eq. (\ref{relationship}) is to extract the elastic neutrino-nucleus cross section from electron-nucleus scattering experimental data (both PC and PV). It is important to stress that in the latter observables the experimental data can be easily reverted to the PWBA results assumed in Eq. (\ref{relationship}) using theoretical models that distort the electron wave function within the nuclear Coulomb field \cite{ruf82}; the same applies to higher-order corrections to the interactions \cite{mus94}. Nevertheless, it is worth discussing the experimental conditions under which Eq. (\ref{relationship}) is best fulfilled. First, Coulomb distortion effects are smaller for more energetic (charged) leptons. Second, as the mass of the target increases, the probability of coherent scattering increases roughly quadratically, and the relative weight of incoherent contributions decreases accordingly; however, nuclear recoil is harder to detect in heavy nuclei, and Coulomb distortion effects are larger (proportional to $Z^2$). Third, low momentum transfers drastically reduce some of the incoherent elastic contributions and reduce the effect of the strangeness content of the nucleon (only applicable to the use of Eq. (\ref{relationship_isoscalar})). 

Since the aim of Eq. (\ref{relationship}) is primarily to obtain neutrino elastic cross sections from electron scattering measurements, it is in order to relate their relative uncertainties:
\begin{equation}\label{relationship_uncertainties1}
\mathcal{E}_{\left(\frac{d\sigma}{d\Omega} \right)_{(\nu,\nu)}}\approx 2 \:\mathcal{E}_{\mathcal{A}_{(e,e)}} \;,
\end{equation}
where the uncertainty in the electron cross section, usually very small, has been neglected. Using this relationship, the relative uncertainty of the neutrino cross section derived from the statistical uncertainty of the PV asymmetry measurement is given by
\begin{equation}\label{relationship_uncertainties}
\mathcal{E}^{stat.}_{\left(\frac{d\sigma}{d\Omega} \right)_{(\nu,\nu)}} \approx 2\:\mathcal{X}_{PV}^{-\frac{1}{2}} \:\mathcal{F}_{PV}^{-\frac{1}{2}} \;,
\end{equation}
where $\mathcal{X}_{PV}$ accounts for the experimental conditions of the PV measurement: total solid angle of the detector, luminosity of the polarized electron beam and running time of the experiment: $\mathcal{X}_{PV}=\Omega \:L \:T$. The figure-of-merit $\mathcal{F}_{PV}$ for a fixed incident electron energy can be expressed in this case as
\begin{equation}\label{fom}
\mathcal{F}_{PV} = \left(\frac{d\sigma}{d\Omega} \right)_{(\nu,\nu)}.
\end{equation}
Thus, by knowing the experimental conditions available and the actual measurements of the PV asymmetry and the PC cross section, one can estimate the precision of the elastic neutrino scattering cross section that can be extracted from Eq. (\ref{relationship}).

To illustrate these ideas we have chosen a set of even-even nuclei that have attracted recent interest in elastic PV electron or neutrino scattering experiments: $^{12}$C, $^{20}$Ne, $^{28}$Si, $^{40}$Ar, $^{48}$Ca, $^{76}$Ge, $^{114}$Cd, $^{130}$Te, $^{132}$Xe, and $^{208}$Pb. For each nuclear target we compute in PWBA the electron (PC) elastic scattering differential cross section and the PV asymmetry in elastic electron scattering using an axially symmetric Skyrme-Hartree-Fock mean field with BCS pairing for the nuclear ground state \cite{vau}. The microscopic calculations used here, although proven successful for a wide variety of processes and in particular for electron scattering off nuclei, are not the main goal of this work; they serve us as substitutes for expected experimental results on PC and PV electron scattering once the Coulomb distortion of the electron has been extracted, and  they are subsequently used to predict the elastic neutrino cross section as in Eq. (\ref{relationship}). Results for the electron scattering (PC or PV) off some of these targets including distortion of the electron wave function can be seen in \cite{mor09_npa,mor10_jpg}.

\begin{figure}
\begin{center}
\includegraphics[trim= 0cm 30cm 0cm 0cm, clip, width=0.45\textwidth]{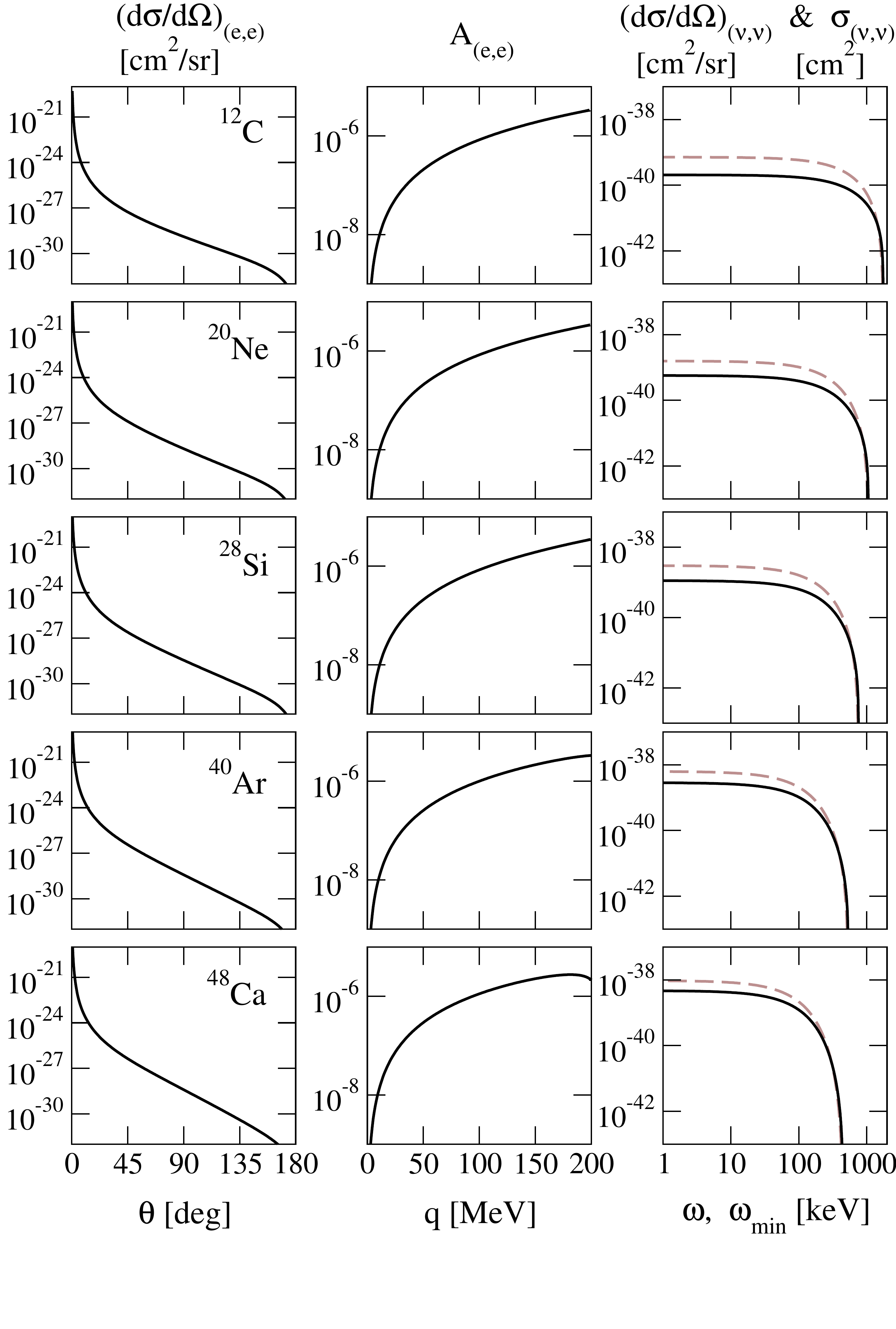}\hspace{1.2cm}
\includegraphics[trim= 0cm 30cm 0cm 0cm, clip, width=0.45\textwidth]{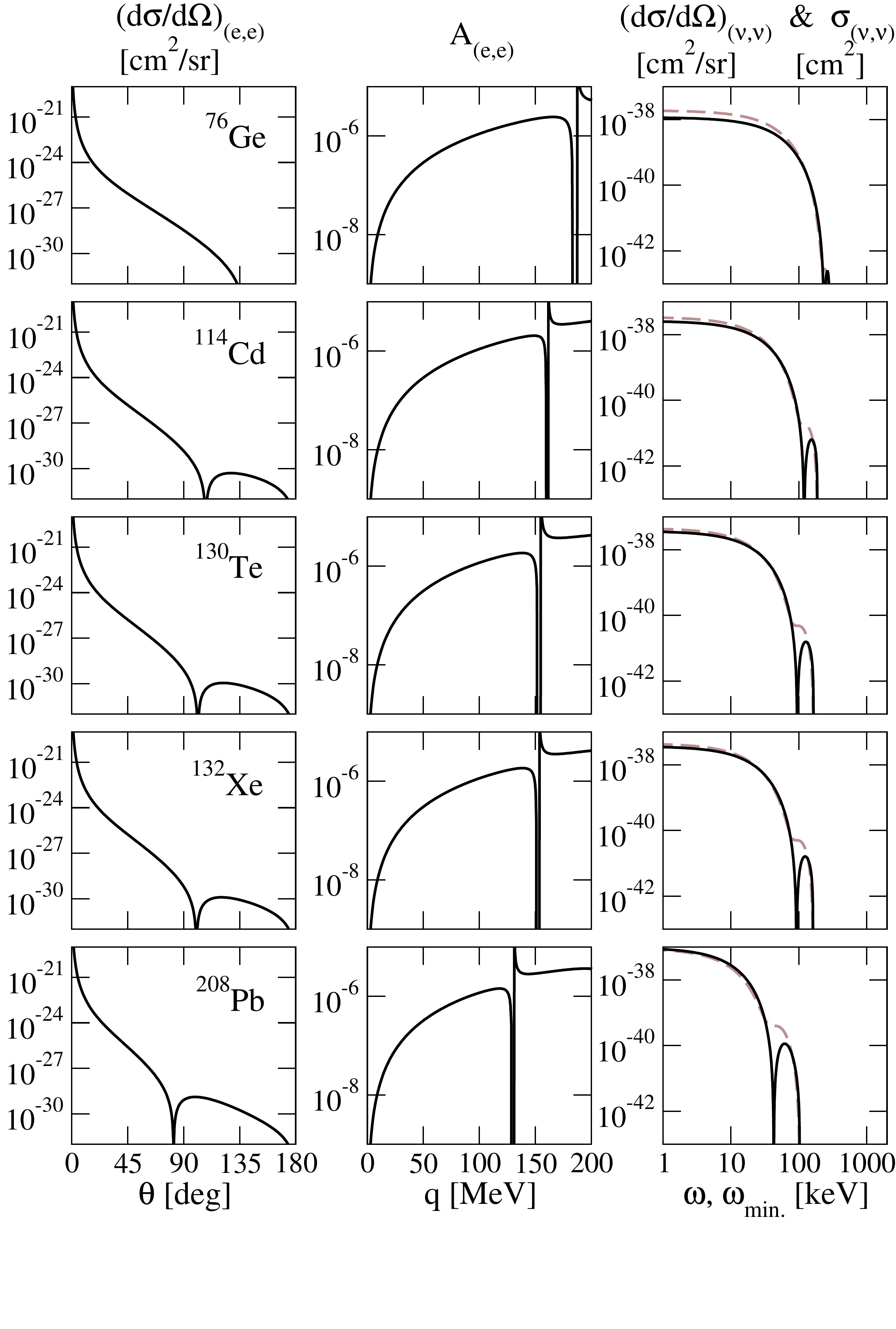}
\caption{For a set of even-even nuclear targets and 100 MeV incident lepton energy, PWBA Skyrme-Hartree-Fock-BCS calculations of electron elastic differential cross sections (left), PV asymmetries in elastic electron scattering (middle), and neutrino elastic differential cross sections (right); in the latter case, dashed curves are cross sections integrated from a minimum detected recoil energy.  \label{elastic}}
\end{center}
\end{figure}

We show our results (solid curves) in Fig. \ref{elastic} for 100 MeV incident lepton energy as a function of a kinematic variable that we consider experimentally suitable in each case: electron elastic scattering differential cross section as a function of the scattering angle (left column), PV asymmetry in elastic electron scattering as a function of the momentum transfer, which is its only kinematic dependence (middle column), and neutrino elastic scattering differential cross section as a function of the recoil energy of the target (right column). The latter are for light active neutrinos of any flavor with $\alpha^{\nu}=1$ (SM value); as per Eq. (\ref{fom}), these curves are also the figures-of-merit at fixed energy of the neutrino cross sections, as well as of the PV asymmetries in the previous column.  In the right column for completeness we also show the integrated neutrino cross sections (dashed curve) as a function of the minimum recoil energy detected ($\omega_{min}$); the abscissa has thus two meanings: running recoil energy for the differential cross sections and minimum recoil energy detected for the integrated cross sections.

In what follows we will show results using neutrinos from the pion decay at rest: $\pi^+ \to \mu^+ + \nu_{\mu}$ (prompt neutrinos), and the subsequent muon decay: $\mu^+ \to e^+ + \nu_{e} + \bar{\nu}_{\mu}$ (neutrinos delayed at the scale of the muon decay lifetime, 2.2 $\mu$s). The spectra of these three types of neutrinos (Michel spectrum) as a function of their energy $\varepsilon$, are given by:
\begin{eqnarray}\label{spectrum1}
&& S_{\bar{\nu}_{\mu}}(\varepsilon) = \frac{16}{m_{\mu}^4} \:\varepsilon^2 \:(3\:m_{\mu}-4\:\varepsilon) \\\label{spectrum2}
&& S_{\nu_{e}}(\varepsilon) = \frac{96}{m_{\mu}^4} \:\varepsilon^2 \:(m_{\mu}-2\:\varepsilon) \\\label{spectrum3}
&& S_{{\nu}_{\mu}}(\varepsilon) = \delta(\varepsilon-\varepsilon_{\pi}), \;\; \text{with} \;\;\varepsilon_{\pi}=\frac{m_{\pi}^2-m_{\mu}^2}{2\:m_{\pi}} \;,
\end{eqnarray}
where $m_{\mu}$ and $m_{\pi}$ are the muon and the pion masses.

\begin{figure}
\begin{center}
\includegraphics[trim= 0cm 24cm 2cm 0cm, clip, width=0.45\textwidth]{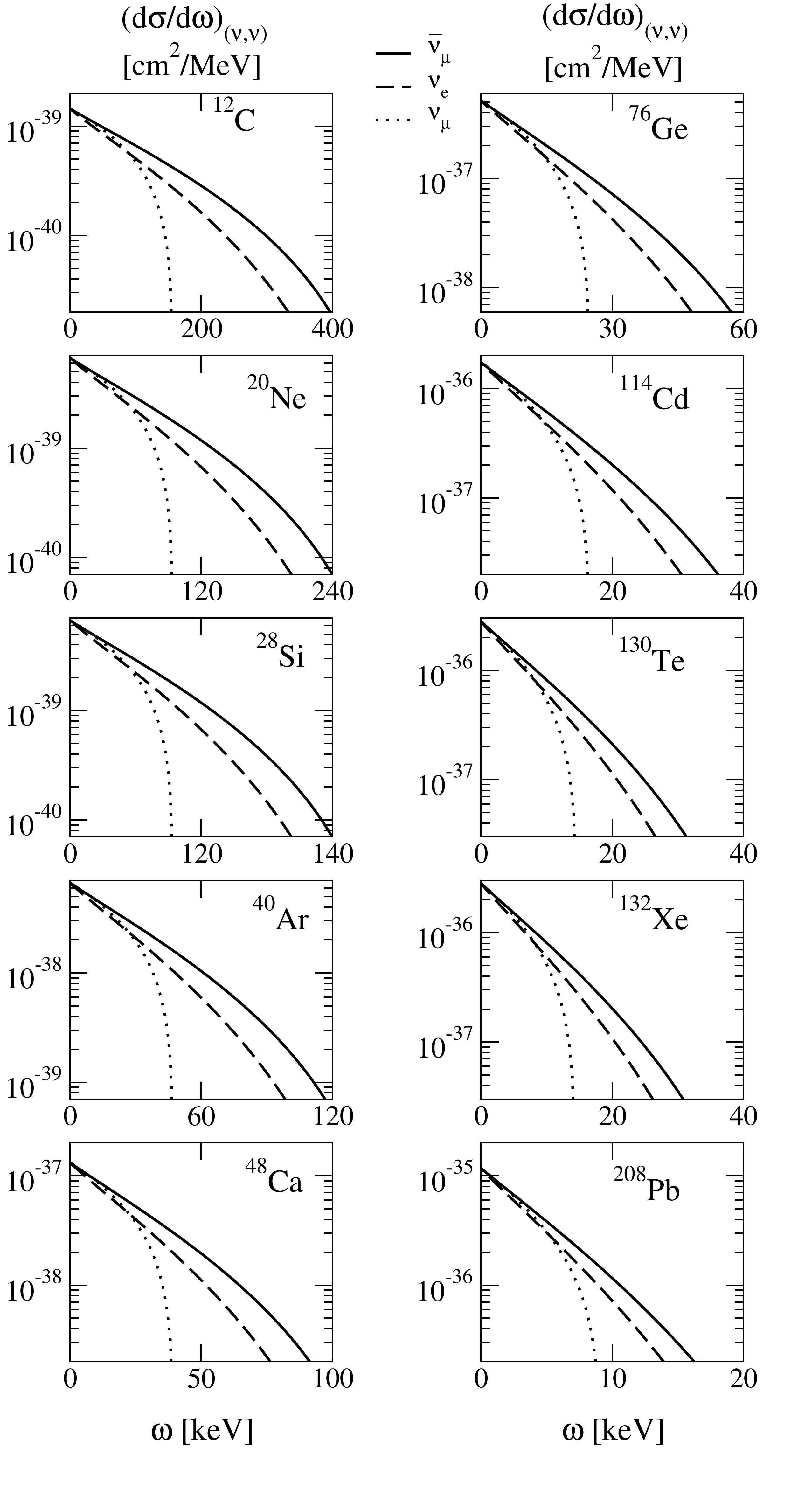}
\caption{For the same set of even-even nuclear targets of Fig. \ref{elastic}, spectrum-folded elastic differential cross sections of the three neutrino types from pion decay at rest: $\bar{\nu}_{\mu}$ (solid curves), ${\nu}_{e}$ (dashed curves), and ${\nu}_{\mu}$ (dotted curves).  \label{elastic_spectra}}
\end{center}
\end{figure}

In Fig. \ref{elastic_spectra} we show, for the same set of even-even nuclear targets of Fig. \ref{elastic}, the elastic neutrino differential cross sections using the three spectra from the pion decay at rest in Eqs. (\ref{spectrum1})-(\ref{spectrum3}). Other calculations of neutrino-nucleus coherent cross-sections in literature, with or without Michel spectrum folding \cite{don82_la,hor03,sch06,and11} show similar results, since the nuclear and nucleon structure details of the target are not very relevant at the low momentum transfers of this process.

In summary, we have provided a relationship, Eq. (\ref{relationship}), to obtain neutrino cross sections from experimental electron scattering data that automatically incorporates the effects of nuclear and nucleon structure details, such as the exact distribution of nucleons or the electric strangeness content of the nucleon (see \cite {mor14_prc} for a detailed analysis of these). Significant deviations of future neutrino data from the results predicted using Eq. (\ref{relationship}) would impact our knowledge of specific properties of neutrinos (not shared by the charged leptons) such as different WNC couplings (in the combination $\alpha^{\nu}$), magnetic moments or the existence of sterile species. On the side of the nuclear target, coherent neutrino scattering has been proposed to probe the neutron density distribution \cite{patton12} as a complement to the use of PV electron scattering. We propose in the next section the study of the axial structure of odd nuclei for which elastic neutrino scattering offers a unique sensitivity.

\section{Axial structure studies with elastic neutrino scattering}

The elastic neutrino cross section is sensitive to the axial structure of the target nucleus through incoherent contributions, whose relative weight is larger under experimental conditions opposite to the ones leading to the validity of Eq. (\ref{relationship}). First, targets must have spin and isospin different from zero, $J\neq 0$, $T\neq 0$, for the axial responses to play a role in the scattering process; even-even nuclei and $N=Z$ nuclei are therefore ruled out. Second, for the axial incoherent contributions not to be overwhelmingly hidden by the coherent part, the latter must be as small as possible; light nuclei are therefore preferred, with the additional advantage of their recoil being larger and therefore easier to detect.

The full WNC matrix element squared of an elastic neutrino-nucleus scattering, as in the cross section of Eq. (\ref{neutrino_xs}), can be decomposed for convenience as follows
\begin{eqnarray}
\widetilde{{\mathcal R}} = \widetilde{{\mathcal R}}_{coh} + \widetilde{{\mathcal R}}_{axial} + \widetilde{{\mathcal R}}_{other} \;,
\label{R}
\end{eqnarray} 
where the coherent term, that has been the main subject of the first part of this paper, is the Coulomb vector isoscalar contribution of Eq. (\ref{R_coh}). The axial term has purely-axial longitudinal and transverse contributions as well as a vector-axial interference contribution:
\begin{eqnarray}
\widetilde{{\mathcal R}}_{axial} = V_{LL} \:(\widetilde{F}^{AA}_{LL})^2 + V_T \:(\widetilde{F}^{AA}_{T})^2 + V_{T'} \:\widetilde{X}^{VA}_{T'}.
\label{R_axial}
\end{eqnarray} 
Finally, an additional incoherent non-axial contribution contains the remaining Coulomb vector contribution and a purely vector transverse contribution:
\begin{eqnarray}
\widetilde{{\mathcal R}}_{other} = V_L \:(\widetilde{F}^{VV}_{CC})^2 - V_L \:(\widetilde{F}^{VV,\:T=0}_{CC,\:J=0})^2 + V_T \:(\widetilde{F}^{VV}_{T})^2
\label{R_other}
\end{eqnarray}
We note that the generalized Rosenbluth factors in the previous expressions contain both purely vector and purely axial terms, $V_X=V_X^{VV}+V_X^{AA}$ (except for the VA interference factor $V_{T'}$), independently of the nature of the accompanying form factor \cite{mor14_prd}.

A convenient observable is the ratio of the part of the differential cross section that is sensitive to the axial current over the full differential cross section:
\begin{eqnarray}
R(q) = \frac{\left(\frac{d\sigma}{d\omega}\right)_{axial}}{\left(\frac{d\sigma}{d\omega}\right)} = \frac{\widetilde{{\mathcal R}}_{axial}}{\widetilde{{\mathcal R}}}.
\label{ratio}
\end{eqnarray}
When the coherent contribution is dominant, and by keeping only the most important contributions in LWL, this ratio can be approximated by
\begin{eqnarray}
R_{LWL} \approx \frac{\widetilde{{\mathcal R}}_{axial}}{\widetilde{{\mathcal R}}_{coh}} \approx \frac{V_T}{V_{L}} \:\frac{(\widetilde{F}^{AA}_{T})^2}{(\widetilde{F}^{VV,\:T=0}_{CC})^2}.
\label{ratio_LWL}
\end{eqnarray}
At $q=0$ it can be estimated as
\begin{eqnarray}\label{ratio_0}
R(0) \approx \frac{\left( \beta^{(1)}_A \:G^{(1)}_A(0)\right)^2}{8\:\sin^4\theta_W} \:\frac{ {\mathcal K}^2_{h.o.}}{(2J+1)\:A^2} \;,
\end{eqnarray}
where $\beta^{(1)}_A$ is the isovector axial WNC coupling ($\beta^{(1)}_A=1$ in SM), $G^{(1)}_A(0)\approx 1.27$ is the value of the isovector axial form factor at $q=0$, $J$ is the nuclear spin, and ${\mathcal K}^2_{h.o.}$ is a factor related to the purely axial transverse form factor $\widetilde{F}^{AA}_{T}(0)$ computed with the odd-nucleon harmonic oscillator wave function within an extreme nuclear shell-model.

We also define a ratio of integrated cross sections,
\begin{eqnarray}
R_{int}(q) = \frac{\sigma_{axial}}{\sigma} \;,
\label{ratio_int}
\end{eqnarray}
which can be used with partially integrated cross sections over the energy acceptance of the recoil energy detector, as in Eq. (\ref{integrated_nu_xs}).

Some a priori suitable target candidates for axial studies whose coherent contribution is dominant are shown in Table \ref{odd_nuclei} with their ground-state spins and isospins, their odd-nucleon harmonic oscillator factor squared ${\mathcal K}^2_{h.o.}$, and the ratio $R$ (in percentage) at $q\to 0$ as in Eq. (\ref{ratio_0}). The relative weight of the axial contribution to the cross-section is larger for $^7$Li (8.6$\%$), $^9$Be (5.2$\%$) and $^{11}$B (3.5$\%$); the expected experimental uncertainties in elastic neutrino scattering, currently around 5$\%$, could be enough to perform axial studies with these targets. As mentioned, the ratios in Table \ref{odd_nuclei} are estimations at $q\to 0$ based on a simple harmonic oscillator shell-model structure; more realistic odd-nucleon wave functions can be built as combinations of different harmonic oscillator states, several of them represented in the table.

\begin{table}[h!]
\caption{\small Selection of even-odd nuclear targets for axial studies with their ground state spin and isospin, odd-nucleon harmonic oscillator factor ${\mathcal K}^2_{h.o.}$ and ratio $R(0)$ in percentage approximated as in Eq. (\ref{ratio_0}). \label{odd_nuclei}} 
\begin{center}
\begin{small}
\begin{tabular}{ccccccc}
\\
\hline
\hline
Isotope \quad & $\quad J^{\pi} \quad$ & $\quad T \quad$ & $\quad {\mathcal K}^2_{h.o.} \quad$ & $R(0)$ [$\%$]  \\
\hline
\\
$^{7}$Li & $3/2^-$ & 1/2 & 4.4 & 8.6    \\
$^{9}$Be & $3/2^-$ & 1/2 & 4.4 & 5.2    \\
$^{11}$B & $3/2^-$ & 1/2 & 4.4 & 3.5    \\
$^{13}$C & $1/2^-$ & 1/2 & 0.4 & 0.3    \\
$^{15}$N & $1/2^-$ & 1/2 & 0.4 & 0.3    \\
$^{17}$O & $5/2^+$ & 1/2 & 5.6 & 1.2   \\
$^{19}$F & $1/2^+$ & 1/2 & 4.0 & 2.1   \\
$^{21}$Ne & $3/2^+$ & 1/2 & 1.6 & 0.3   \\
$^{23}$Na & $3/2^+$ & 1/2 & 1.6 & 0.3   \\
$^{25}$Mg & $5/2^+$ & 1/2 & 5.6 & 0.6   \\
$^{27}$Al & $5/2^+$ & 1/2 & 5.6 & 0.5   \\
\end{tabular}
\end{small}
\end{center}
\end{table}

\begin{figure}
\begin{center}
\includegraphics[width=0.48\textwidth]{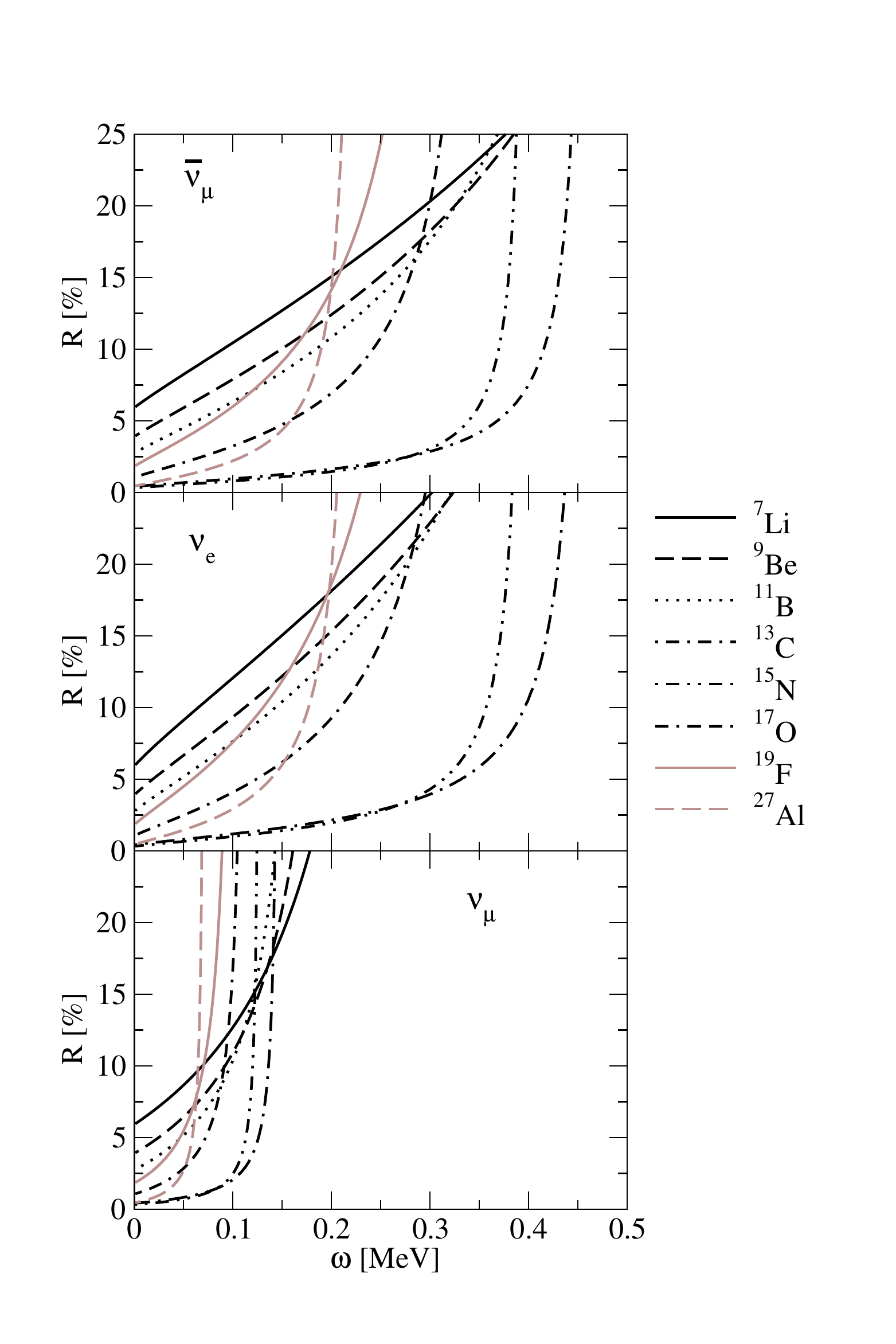}
\caption{Ratio $R$ of axial contributions to the spectrum-folded differential cross section for a set of a priori suitable candidates, using the $\bar{\nu}_{\mu}$ (upper plot), ${\nu}_{e}$ (middle plot) and ${\nu}_{\mu}$ (lower plot) spectra from pion decay at rest. \label{comparison_R}}
\end{center}
\end{figure}

Figure \ref{comparison_R} shows the ratio $R$ of axial contributions to the full cross section, as in Eq. (\ref{ratio}), for some of the nuclei in Table \ref{odd_nuclei}, using the $\bar{\nu}_{\mu}$ (upper plot), ${\nu}_{e}$ (middle plot) and ${\nu}_{\mu}$ (lower plot) spectra from pion decay at rest. Incoherent axial contributions used in these ratios have been estimated using the odd-nucleon harmonic oscillator wave function of an extreme shell model.

\begin{figure}
\begin{center}
\includegraphics[width=0.45\textwidth]{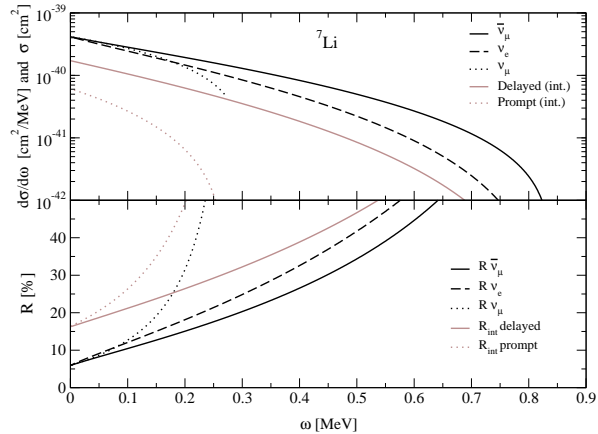}
\caption{Upper panel:  Elastic neutrino cross sections off $^7$Li differential with respect to the recoil energy for pion decay-at-rest neutrino spectra as a function of the recoil energy (dark curves) and integrated cross section for delayed and prompt neutrinos from the same process as a function of the minimum detected recoil energy (light curves). Lower panel: Corresponding ratios $R$ and $R_{int}$ (in percentage) of the axial contribution over the full cross section. \label{xs_and_ratio_li7}}
\end{center}
\end{figure}

\begin{figure}
\begin{center}
\includegraphics[width=0.45\textwidth]{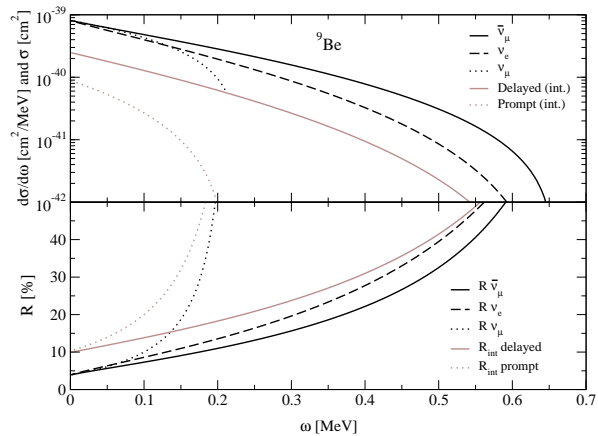}
\caption{As for Fig. \ref{xs_and_ratio_li7}, but for a $^9$Be target.  \label{xs_and_ratio_be9}}
\end{center}
\end{figure}

This comparison shows how the lighter nuclei, $^{7}$Li, $^{9}$Be and $^{11}$B, have the largest ratios in the region of low recoil energy. For larger recoil energies the full cross section decreases, which considerably reduces the statistics of the measurement. To show this effect we plot in Figs. \ref{xs_and_ratio_li7} and \ref{xs_and_ratio_be9}, for the most favorable cases $^7$Li and $^9$Be, the spectrum-folded differential $d\sigma/d\omega$ and integrated $\sigma$ cross sections (upper panel) together with the axial vs. full ratio for each case, $R$ and $R_{int}$ respectively (lower panel) as a function of the energy transfer $\omega$, again for the three neutrino spectra from pion decay at rest. The integrated cross sections and the corresponding ratios are obtained upon integration over the acceptance window of recoil energy from a given minimum value, the detector threshold, to the maximum value kinematically allowed (Eq. (\ref{integrated_nu_xs})); those curves are shown separately for delayed neutrinos ($\bar{\nu}_{\mu}$ and $\nu_{e}$) and for prompt neutrinos ($\nu_{\mu}$).

\begin{figure}
\begin{center}
\includegraphics[width=0.45\textwidth]{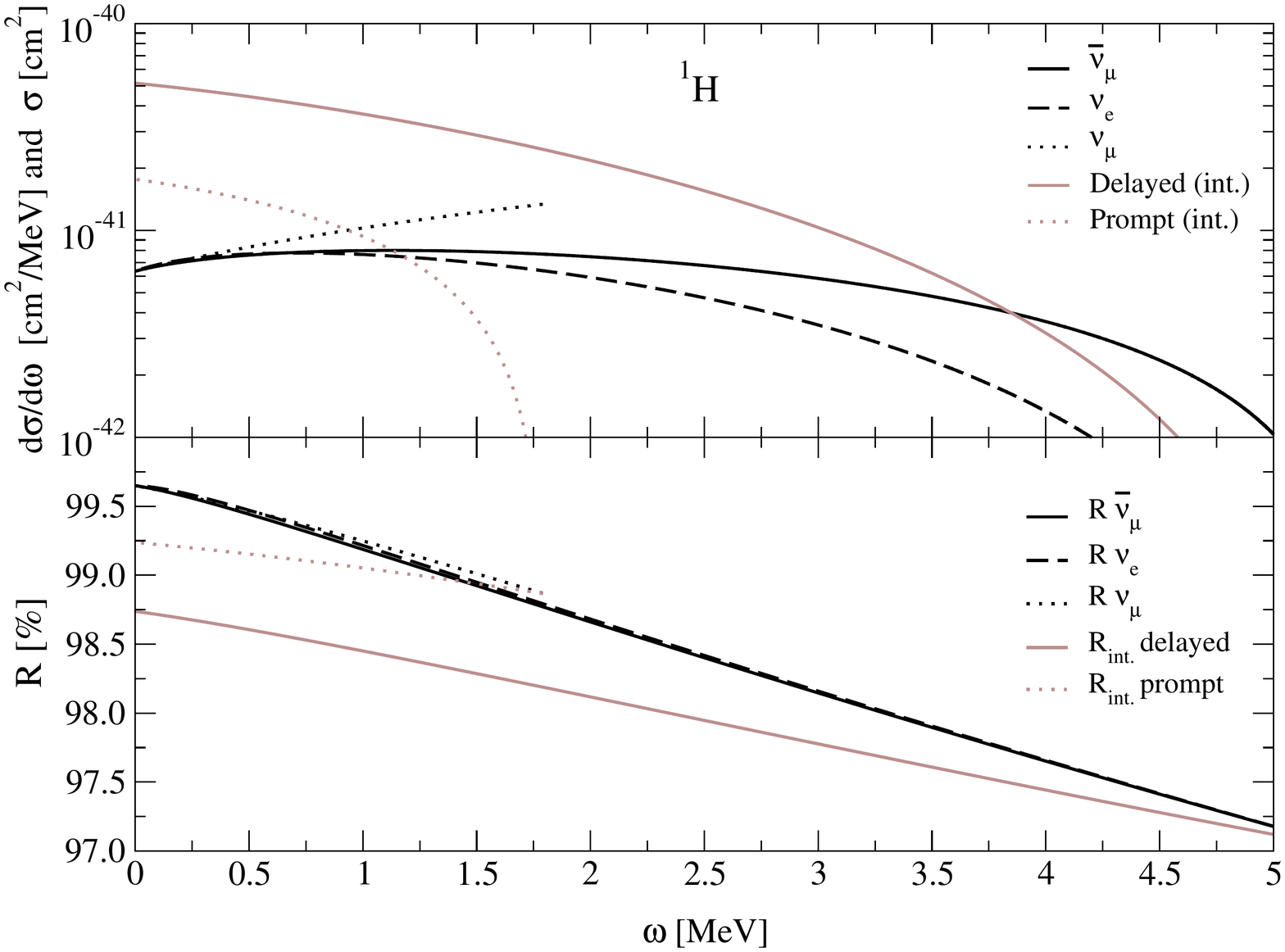}
\caption{As for Fig. \ref{xs_and_ratio_li7}, but for $^1$H target.  \label{xs_and_ratio_h1}}
\end{center}
\end{figure}

\begin{figure}
\begin{center}
\includegraphics[width=0.45\textwidth]{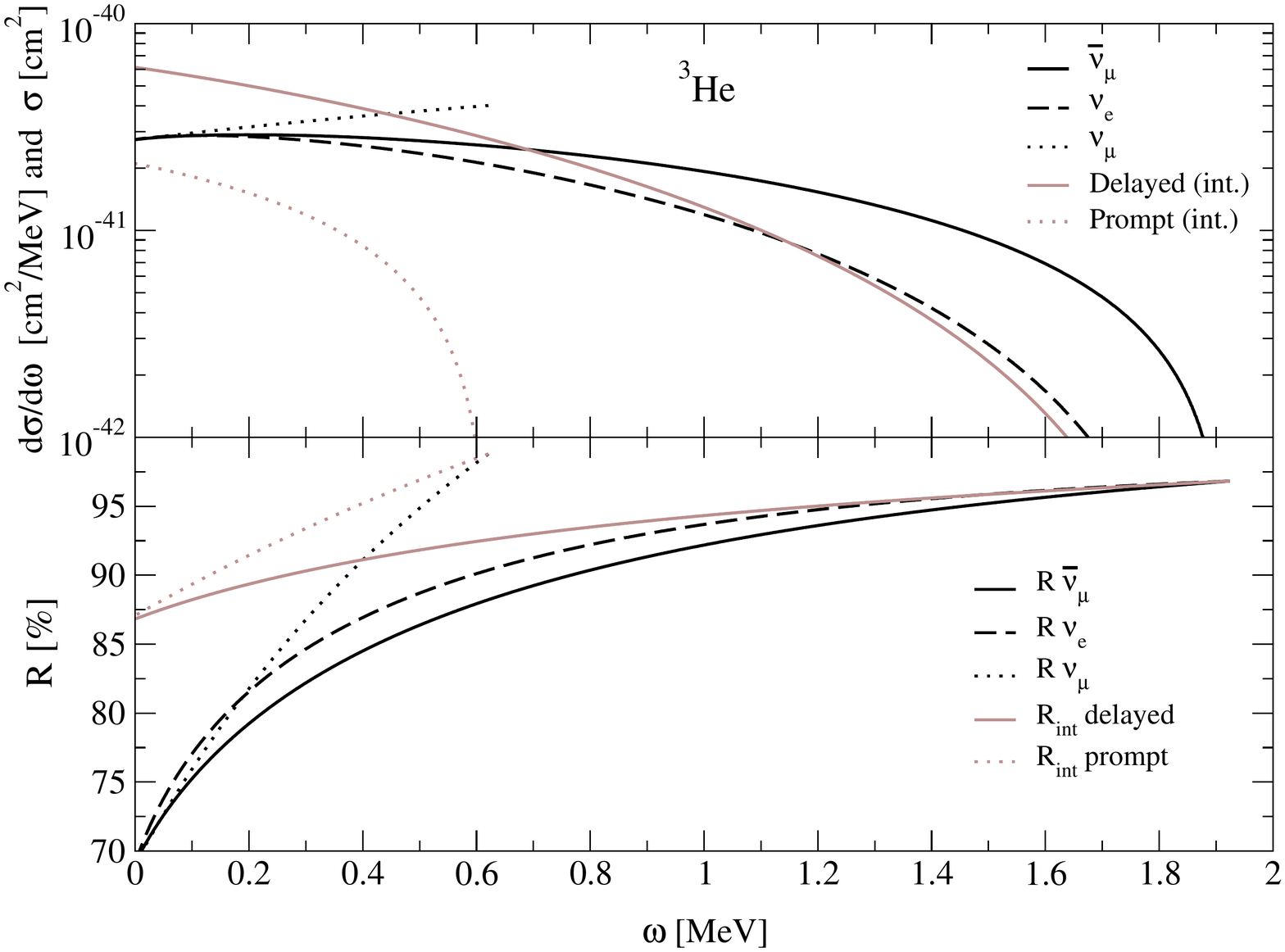}
\caption{As for Fig. \ref{xs_and_ratio_li7}, but for a $^3$He target.  \label{xs_and_ratio_he3}}
\end{center}
\end{figure}

For very light odd targets with non-dominant coherent contribution, $^1$H (proton) and $^3$He, we show cross sections and axial-over-total ratios in Figs. \ref{xs_and_ratio_h1} and \ref{xs_and_ratio_he3}. The ratios are much larger than in the cases analyzed above, showing a clear dominance of the axial contribution due to the smallness of the coherent one. The cross sections are, however, smaller than in the cases above, due again to the small coherent enhancement; this fact reduces the statistics and therefore the suitability of these nuclei for precision studies.

O. M. acknowledges support from a Marie Curie International Outgoing Fellowship within the 7th European Community Framework Programme (ELECTROWEAK project). Also supported in part by the Office of Nuclear Physics of the US Department of Energy under Grant Contract Number DE-FG02-94ER40818 (T. W. D.).

\end{document}